\documentstyle[12pt]{article}
\def\beq{\begin{equation}}
\def\eeq{\end{equation}}
\def\bea{\begin{eqnarray}}
\def\eea{\end{eqnarray}}
\def\bq{\begin{quote}}    
\def\eq{\end{quote}}      
\def\ra{\rightarrow}

\def\bq{\begin{quote}}
\def\eq{\end{quote}}  
\def\ra{\rightarrow}

\begin{document}

\baselineskip 24pt
\newcommand{\sheptitle}
{Minimal Supersymmetric
$\bf SU(4) \times SU(2)_L \times SU(2)_R$}

\newcommand{\shepauthor}
{S. F. King$^{\dagger}$
\footnote{On leave of absence from 
Department of Physics and Astronomy,
University of Southampton, Southampton, SO17 1BJ, U.K.}
and Q. Shafi$^\ast $}

\newcommand{\shepaddress}
{$^\dagger$ Theory Division, CERN, CH-1211 Geneva 23, Switzerland\\
$^\ast$Bartol Research Institute, University of Delaware\\
Newark, DE 19716, USA}

\newcommand{\shepabstract}
{We present a minimal string-inspired supersymmetric 
$SU(4) \times SU(2)_L \times SU(2)_R$ model, and provide a
detailed analysis of the symmetry breaking potential in this model, 
based on a generalisation of that recently proposed 
by Dvali, Lazarides and Shafi.
The model contains a global $U(1)$ R-symmetry and reduces to the
MSSM at low energies. However it improves on the MSSM since it
explains the magnitude of its $\mu$ term and gives a prediction for
$\tan \beta \simeq m_t/m_b$. It also 
predicts an essentially stable proton, and 
contains both `cold' and `hot' dark matter candidates. 
A period of hybrid inflation
above the symmetry breaking scale is also possible in this model.
Finally it suggests the 
existence of `heavy' charge $\pm e/6$ (colored) and $\pm e/2$ (color 
singlet) states.  }

\begin{titlepage}
\begin{flushright}
CERN-TH/97-314\\
BA-97-49\\
hep-ph/9711288\\
\end{flushright}
\begin{center}
{\large{\bf \sheptitle}}
\bigskip \\ \shepauthor \\ \mbox{} \\ {\it \shepaddress} \\ \vspace{.5in}
{\bf Abstract} \bigskip \end{center} \setcounter{page}{0}
\shepabstract
\begin{flushleft}
CERN-TH/97-314\\
\today
\end{flushleft}
\end{titlepage}

In a recent paper by Dvali, Lazarides and Shafi (DLS)
\cite{dls}, the electroweak 
sector of the minimal supersymmetric standard model (MSSM) was embedded 
in the supersymmetric gauge group $SU(2)_L \times SU(2)_R \times 
U(1)_{B-L}$.  \footnote{Original references to this gauge
group, which is not our main concern here, may be found in \cite{dls}.}
By imposing a global R-symmetry, $U(1)_R$,
which also happens to 
contain the unbroken MSSM R-parity as a $Z_2$ subgroup, 
it was shown in \cite{dls} that 
the `low energy' limit of this extension gives rise to a number of 
interesting consequences.  For instance, one could relate the magnitude 
of the supersymmetric $\mu$-term of MSSM to the supersymmetry (SUSY) 
breaking scale ($m_{3/2}$), and also automatically generate the B$\mu$ 
term.  The apparent stability of the proton ($\tau_p > 10^{32} - 
10^{33}$ yrs), which is somewhat challenging for MSSM since dimension 
five proton decay is in principle allowed, is explained to be a 
consequence of an accidental global $U(1)_B$ symmetry of the underlying 
superpotential.  Finally, as one would expect, the extended gauge 
symmetry gives rise to non-zero neutrino masses which may play a role in 
the solar neutrino puzzle, dark matter issues, and so on.

In this note we 
extend the DLS approach to the Pati-Salam gauge group \cite{pati}
\begin{equation}
G_{PS} \equiv SU(4) \times SU(2)_L \times SU(2)_R
\end{equation}
Although the generalisation is essentially straightforward, we consider
such an extension worthwhile for two reasons.
The first is that $G_{PS}$ can arise from certain 
four dimensional superstring constructions, which makes it an attractive 
candidate for the effective theory below the string scale \cite{alr}.  
The second reason is that although there has been a good deal of 
attention focussed on such a string-inspired model \cite{422}
there has been very little detailed 
discussion of the symmetry breaking mechanism
which must be present in this model. In fact we are unaware
of any analysis of the symmetry breaking potential of the string-inspired
version of this model in which only Higgs in fundamental representations
are permitted. The DLS mechanism will be shown to be ideally suited
for this purpose, and may be extended in a completely trivial way
to the higher gauge symmetry. Apart from breaking the gauge symmetry
correctly, the DLS mechanism also
solves the $\mu$ problem, and provides a mechanism for hybrid
inflation.

The resulting scheme may be regarded as minimal in the sense that the 
higgs sector contains the minimum number of fields needed to implement 
the gauge symmetry breaking.  As in \cite{dls}, the model may be supplemented 
by a $U(1)_R$ symmetry which contains the MSSM $Z_2$ R-parity.  We 
show how our approach leads to a more robust MSSM at low energies.  
Namely, why the magnitude of the $\mu$-term is $\sim TeV$, why the 
lightest (CP even) scalar has not been seen at LEP II, and why/how the 
proton happens to be so stable.  Moreover, since $\tan \beta \simeq 
m_t/m_b$, one is able to be quite a bit more specific about the 
sparticle spectrum, as well as the composition of the LSP (essentially a 
`bino').  Finally, the model contains both `cold' and `hot' dark matter 
candidates, which seems desirable from studies of large scale structure 
formation \cite{ss}.

It has been realised for some time that
the Pati-Salam gauge symmetry may be broken at the scale $M \sim M_{GUT}$
by the following Higgs representations which arise naturally in the
superstring construction \cite{alr}, 
\bea
{\bar{H}} & = & (\bar{4},1,\bar{2})= (u^c_H,d^c_H,\nu^c_H,e^c_H ) \nonumber \\
{H} & = & (4,1,2) = (\bar{u}^c_H,\bar{d}^c_H,\bar{\nu}^c_H,\bar{e}^c_H )
\eea
The neutral components of the Higgs fields were 
assumed to develop VEVs
\begin{equation} <\tilde{\nu_H^c}>=<{\tilde{\bar{\nu_H^c}}}>\sim M,
\label{HVEV}
\end{equation} leading to the symmetry breaking at $M$:
\begin{equation}
G_{PS}\rightarrow
\mbox{SU(3)}_C \otimes \mbox{SU(2)}_L \otimes \mbox{U(1)}_Y
\label{422to321}
\end{equation} in the usual notation.
The quarks and leptons are unified in the representations
$$
F_i = (4,2,1)_i = \left(\begin{array}{cccc}u&u&u&\nu\\
d&d&d&e\end{array}\right)_i
$$
\beq
\bar{F}_i = (\bar{4},1,\bar{2})_i = 
\left(\begin{array}{cccc}u^c&u^c&u^c&\nu^c\\
d^c&d^c&d^c&e^c\end{array}\right)_i
\eeq
\noindent where the subscript $i\ (=1,2,3)$ denotes the family index.
The two low energy Higgs doublets of the MSSM
are contained in the following representation,
\begin{equation}
h=(1,2,\bar{2})=
\left(\begin{array}{cc}
{h_1}^0 & {h_2}^+ \\   
{h_1}^- & {h_2}^0      
\end{array} \right) \label{h}
\end{equation}

We now introduce the following superpotential
\bea
W & = & S[\kappa ({\bar{H}} H  - M^2) +  \lambda  h^2]
+ \lambda_6D_6HH + \bar{\lambda}_6D_6\bar{H}\bar{H} \nonumber \\
& + &  g_{33} \bar{F}_3 F_3 h 
+\ g_{ij} \bar{F}_iF_j h\frac{(\bar{H} H)^n}{M_P^{2n}} 
+ \ h_{ij} \frac{\bar{F}_i \bar{F}_j H H}{M_P} 
\label{superpotential}
\eea
where $S$ denotes a $G_{PS}$ singlet superfield, the parameters $\kappa, 
\lambda$ and $M$ are taken to be real and positive, and $h^2$ denotes 
the unique bilinear invariant $\epsilon^{ij} h^{(1)}_i h^{(2)}_j$.  
Also, $M_P (\simeq 2.4 \times 10^{18}$ GeV) denotes the `reduced' Planck 
mass.  

The superpotential $W$ in Eq.\ref{superpotential} carries two units of charge
under an assumed global $U(1)_R$ symmetry, where $R_{\theta}=1$ leads
to an R-invariant contribution to the potential $V = \int d^2\theta W$.
The R-charges of the various superfields are assigned as follows: 
\beq
R_{D_6}=R_S = 2, 
\ R_H = R_{\bar {H}} = R_h = 0, \ R_F = R_{\bar{F}} = 1.
\eeq
This $U(1)_R$ symmetry closely resembles that proposed for the MSSM
by Hall and Randall \cite{HR}, and contains a $Z_2$ subgroup which may
be identified with R-parity.
\footnote{An alternative R-symmetry in which the MSSM Higgs superfields
have unit R-charges, while the matter superfields have half-integer
R-charge contains a $Z_2$ subgroup which may be identified with 
`matter parity' \cite{Dim}.
However such a choice of R-charges would allow the $\mu$ term
whereas our choice of R-charges forbids it.}
The usual problem with such global R-symmetries is that they are
violated by the terms in the soft supersymmetry breaking potential,
$V_{soft}$ and in particular the gaugino masses, since the gauginos
carry one unit of R-charge. Here we consider a scenario in which
the superpotential in the visible sector respects the $U(1)_R$ symmetry,
while the hidden sector also respects the R-symmetry, but spontaneously
breaks it leading to an R-violating $V_{soft}$. 
\footnote{The would-be Goldstone boson may receive a mass from the
non-perturbative sector of the hidden sector.
Note that terms in the visible superpotential which
are forbidden by R-symmetry, may alternatively be forbidden by suitable
discrete symmetries.}

As discussed in \cite{alr}
the high energy Higgs mechanism removes the $H,\bar{H}$ components
$u^c_H,e^c_H,\bar{u}^c_H,\bar{e}^c_H$ from the physical spectrum  
(half of these states get eaten by the heavy gauge bosons and gauginos
and the other half will become massive Higgs bosons),
leaving massless $d^c_H,\bar{d}^c_H$. In order to give these
states a large mass, following \cite{alr}
we have introduced a colour sextet
superfield
$$
D_6=(6,1,1)=(D^c,\bar{D^c}),
$$
where
$D^c=(\bar{3},1,\frac 13)$ and
$\bar{D}^c=(3,1,- \frac 13)$. 
The way the colour triplets receive
superheavy masses is the following.
Remember first that the decomposition of the sextet gives an
antitriplet/triplet pair
($D_6\ra D^c+\bar{D^c}$). On the other hand
$\bar{H}, H$ fields contain also another such pair with the
same quantum numbers: $d^c_H, \bar{d}_H^c$.
To break the $SU(4)\times SU(2)_R$
the $H,\bar{H}$ fields acquire VEVs of ${\cal O}(M)$. Then 
one  gets the
following two mass terms
 \beq
 <H> H D_6 +<\bar{H}>\bar{H} D_6\ra <\tilde{\bar{\nu}}_H>\bar{d}^c_HD^c+
 <\tilde\nu_H^c>d_H^c \bar{D^c}
 \label{Dmass}
 \eeq
so that the `low energy' limit of the 
theory is precisely MSSM.  

Note that the R-symmetry prevents the 
occurence of the gauge allowed couplings $D_6FF,\ D_6 {\bar{F}} {\bar{F}}$.  
This means that the Higgs colour triplet mediated (dimension five and six) 
baryon number violating processes are entirely eliminated.
Since there is no gauge-mediated proton decay in this model,
this means that the proton appears to be completely stable in the
presence of our assumed $U(1)_R$. However, as noted, this symmetry must
be broken in the hidden sector of the theory, which leaves open the possibility
of baryon number violation arising from the hidden sector, although one
would expect it to be quite suppressed relative to the dimension five terms.

As regards the fermion mass operators, following \cite{K} we have
assumed that only the third family receives a renormalisable 
contribution to its mass, while the other entries of the Yukawa matrix
are filled out by non-renormalisable operators of the type shown,
including an operator which will contribute to large Majorana
masses for the right-handed neutrinos.
This leads to third family Yukawa unification \cite{masses0}, and
predictions for 
light quark and lepton masses and mixing angles \cite{masses1}. 
This scheme may be enforced by introducing a further gauged $U(1)_X$ family
symmetry \cite{masses2}, although we have not done so here.

For our present purposes, the superpotential terms of greatest interest
are those involving the gauge singlet field $S$. Thus we shall consider
the superpotential:
\beq
W_S  =  S[\kappa ({\bar{H}} H  - M^2) +  \lambda  h^2]
\eeq
which leads to the potential $V=V_F+V_D+V_{soft}$ where
\beq
V_F  =  
|\kappa (\bar{H}H-M^2)+\lambda h^2|^2
+|\kappa S \bar{H}|^2 +|\kappa S H|^2 +|\lambda S h|^2
\eeq
\beq
V_{soft}=m^2(|S|^2+|\bar{H}|^2+|H|^2+|h|^2)
+(A_{\kappa}S\bar{H}H+A_{\lambda}Sh^2-A_1SM^2 +H.c.)
\eeq
We take the dimensionless couplings $\lambda$ and $\kappa$ to be real
and positive, and the mass parameter $M$ to be real and of order $10^{16}$ GeV.
The minimal K\"{a}hler potential considered in \cite{dls} corresponds
to the special choice $m=m_{3/2}$, $A_{\lambda}=\lambda Am_{3/2}$,
$A_{\kappa}=\kappa Am_{3/2}$, $A_1=\kappa (A-2)m_{3/2}$. 
\footnote{For simplicity we have taken the
soft masses to be all equal even in the non-minimal case, as this assumption
does not qualitatively change the results.}
As in \cite{dls} we shall impose the D-flatness conditions, on the
neutral components of the fields which corresponds to
$|h_1^0|=|h_2^0|\equiv h$,
$|\tilde{\nu_H^c}|=|{\tilde{\bar{\nu_H^c}}}|\equiv \nu$,
with all charged and coloured components set equal to zero.

Neglecting the phases, and writing $s=|S|$,
the potential reduces to
\bea
V & = & 
[\kappa (\nu^2-M^2)+\lambda h^2]^2+2\kappa^2 s^2 \nu^2 +2\lambda^2 s^2h^2
\nonumber \\
& + & 2m^2(s^2/2+\nu^2+h^2)+2A_{\kappa}s\nu^2+2A_{\lambda}sh^2-2A_1sM^2
\eea
In the absence of $V_{soft}$ terms the potential
is easily seen to be minimised by $s=0$ with $\nu$ and $h$ lying somewhere 
on a flat direction characterised by
the ellipse $F_s\equiv [\kappa (\nu^2-M^2)+\lambda h^2]=0$.
When $V_{soft}$ terms are
included, the flat direction $F_s=0$ characterised by 
the ellipse will be lifted
and a particular VEV for $h$ and $\nu$ will be selected.
Including the soft terms and equating $\frac{\partial V}{\partial s}=0$
we obtain the $s$ VEV in terms of the other VEVs
\beq
s=\frac{-(A_{\kappa}\nu^2+A_{\lambda}h^2-A_1M^2)}{2(\kappa^2\nu^2+\lambda^2h^2)
+m^2}
\eeq
which, for any choice of $\nu$ and $h$ on or near the ellipse, corresponds to a
small VEV of $s\sim m_{3/2}$ which will lead to ``effective''
$\mu$ and $B\mu$ terms of the correct order of magnitude.
The condition $\frac{\partial V}{\partial \nu}=0$ is satisfied by
either $\nu=0$ or $F_s=-(\kappa^2s^2+A_{\kappa}s+m^2)/\kappa$,
while the condition $\frac{\partial V}{\partial h}=0$ is satisfied by
either $h=0$ or $F_s=-(\lambda^2s^2+A_{\lambda}s+m^2)/\lambda$.
There are thus two natural candidate minima: the ``good'' point
with $h=0$ and the ``bad'' point with $\nu=0$. Both cases correspond to a
violation of the ellipse condition $F_s=0$ by an amount of order $m_{3/2}^2$.
However this does not mean that the term $F_s^2\sim m_{3/2}^2$ may be
neglected in the analysis of the potential compared to other terms of order
$m_{3/2}^2M^2$, since as we shall see the question of the stability
around each extremum is a delicate matter for this potential.

By expanding the minimisation conditions for
the ``good'' point $h=0$, we find that
$s\approx -(A_{\kappa}-A_1)/(2\kappa^2)+\delta s$, 
$\nu^2\approx M^2 -\delta \nu^2$
where $\delta \nu^2=(4\kappa^2s^2+2m^2+4A_{\kappa}s)/(4\kappa^2)$,
and $\delta s =(-2m^2s+2A_1\delta \nu^2)/(4\kappa^2M^2)$,
where the lowest order VEVs may be consistently inserted into the 
corrections $\delta \nu^2$, $\delta s$. The stability of the ``good''
case $h=0$ is governed by the discriminant consisting of the three
dimensional determinant of second derivatives. However in this case
the discriminant turns out to be simply proportional to the Higgs mass
\beq
\frac{\partial^2V}{\partial h^2}=4\lambda F_s+4\lambda^2s^2+4A_{\lambda}s+4m^2
\eeq
Inserting the expansions for the VEVs explicitly we find
\bea
\frac{\partial^2V}{\partial h^2} & = & (\lambda^2A_1^2-\lambda \kappa A_1^2
-2\lambda^2A_{\kappa}A_1+\lambda \kappa A_{\kappa}^2 + \lambda^2A_{\kappa}^2
\nonumber \\
& - & 2\kappa^2A_{\lambda}A_{\kappa}
+2\kappa^2A_{\lambda}A_1
+4\kappa^4m^2-2\lambda \kappa^3m^2)/(\kappa^4)
+\cdots
\eea
The height of the potential at the ``good'' point is
\beq
V(h=0)=(4\kappa^2m^2+2A_{\kappa}A_1-A_{\kappa}^2-A_1^2)M^2/(2\kappa^2)+\cdots
\eeq
The sign of the Higgs mass squared
(whose magnitude
is of order $m_{3/2}^2$ which implies that stability is a delicate matter)
governs the stability of case $h=0$.
For example in the minimal K\"{a}hler case we find 
$\frac{\partial^2V}{\partial h^2}
=2m_{3/2}^2(2(\frac{\lambda}{\kappa})^2 -3(\frac{\lambda}{\kappa})+2 )$
which is always positive indicating that the ``good'' case
$h=0$ is always a local minimum. The height of the potential in this
case is $V(h=0)=-\frac{m_{3/2}^4}{8\kappa^2}(8A^2-32A+22)$, corresponding to
a remarkable cancellation of the leading
terms of order $m_{3/2}^2M^2$ at this point.
In the non-minimal case, the stability is still governed by a Higgs 
mass of the same order but now it may or may not correspond to a local
minimum depending on choice of soft parameters,
while the height of the potential is of order $m^2M^2$.

Turning now to the ``bad'' point $\nu=0$ we find the corresponding
results $s\approx -(A_{\lambda}-A_1\frac{\lambda}{\kappa})/(2\lambda^2)
+\delta s$, 
$h^2\approx \frac{\kappa}{\lambda}M^2 -\delta h^2$
where $\delta h^2=(4\lambda^2s^2+2m^2+4A_{\lambda}s)/(4\lambda^2)$,
and $\delta s =(-2m^2\frac{\kappa}{\lambda}s+2A_1\delta h^2)/(4\kappa^2M^2)$,
where the lowest order VEVs may be consistently inserted into the 
corrections $\delta h^2$, $\delta s$. The stability of the ``bad''
point $\nu =0$ is governed by the $\nu$ mass
\beq
\frac{\partial^2V}{\partial \nu^2}
=4\kappa F_s+4\kappa^2s^2+4A_{\kappa}s+2m^2
\eeq
\bea
\frac{\partial^2V}{\partial \nu^2} & = & ( \lambda^2\kappa A_1^2-\lambda^3A_1^2
+2\lambda^3A_{\kappa}A_1+\kappa^3 A_{\lambda}^2 + \lambda \kappa^2A_{\lambda}^2
\nonumber \\
& - & 2\lambda^2\kappa A_{\lambda}A_{\kappa}
-2\lambda \kappa^2A_{\lambda}A_1
+4\lambda^4\kappa m^2-2\lambda^3 \kappa^2m^2 )/(\kappa \lambda^4)
+\cdots
\eea
The height of the ``bad'' point is
\beq
V(\nu=0)=
(4\lambda^2\kappa^2m^2
+2\lambda \kappa A_1A_{\lambda}
-\kappa^2A_{\lambda}^2-\lambda^2A_1^2)M^2/(2\lambda^3\kappa )
+\cdots 
\eeq
For example in the minimal K\"{a}hler case we find 
$\frac{\partial^2V}{\partial \nu^2}
=2m_{3/2}^2(2(\frac{\kappa}{\lambda})^2-3(\frac{\kappa}{\lambda})+2 )$
which again is always positive, indicating that the ``bad'' point
$\nu=0$ is also always a local minimum. The height of the potential in this
case is $V(\nu=0)=-\frac{m_{3/2}^4}{8\lambda^2}(8A^2-32A+22)$, 
again corresponding to a cancellation of leading terms. 

To summarise, in the minimal case the ``good'' and ``bad'' minima are 
located in very deep wells corresponding to very accurate cancellation
of large terms, with the wells separated by a large potential
barrier of order $m_{3/2}^2M^2$. In the non-minimal case 
the accurate cancellation does not take place, and the ``good'' and ``bad''
points may or may not correspond to local minima, depending on the 
particular values of the parameters.
Several comments are in order:
\begin{description}
\item[i.] At tree level, and in the absence of SUSY breaking, the vacuum 
corresponds to 
$$
<S> = 0, \ \kappa {\bar{H}} H + \lambda h^2 = \kappa M^2
$$
\beq
H = e^{i \phi} \bar{H}^*,\ h^{(1)}_i = e^{i \theta} \epsilon_{ij} 
h^{(2)j*}
\eeq 
\item[ii.] After including SUSY breaking terms \'{a} la minimal 
supergravity (SUGRA), the $S$ field acquires a VEV
given by $<S>\ = - m_{3/2}/\kappa$.  This generates the MSSM 
parameter $\mu = \lambda <S> = -\frac{\lambda}{\kappa} m_{3/2}$.  
Moreover, it turns out that \cite{dls}
\beq
B \mu \simeq -\frac{2 \lambda}{\kappa} m^2_{3/2}
\eeq
\item[iii.] With $\lambda \neq \kappa$, and with minimal SUGRA, there 
are two local minima of the potential, namely the `desired' one $h=0,\ 
H={\bar{H}}=M(\sim M_{GUT})$, and the `bad' one with $h \neq 0, \ H = 
{\bar{H}} = 0$. As we have discussed in detail,
the latter can be eliminated by invoking a non-minimal 
K\"{a}hler potential such that only the `desired' minimum survives.
\item[iv.] With $<H> \sim M_{GUT}$, the heaviest right-handed neutrino 
mass is on the order of $M^2_{GUT}/M_P \simeq 10^{13} - 10^{14}$ GeV.  
This follows from the last non-renormalizable term in 
Eq.\ref{superpotential}.  
Coupled with the fact that $SU(4)$ implies the asymptotic relation 
$m^{(0)}_{\nu_{\tau}} ({\rm Dirac}) = m^{(0)}_{\tau}$, this suggests 
that the `tau' neutrino may, via the usual see-saw mechanism, acquire mass in 
the eV range. In addition it has been shown that the `electron' neutrino
and `muon' neutrino typically have masses and mixing angles in the range
suitable for the MSW solution to the solar neutrino problem,
with `muon-tau' neutrino mixing angles in the observable range of the
CHORUS experiment \cite{masses1}.
\item[v.] The $Z_2$ subgroup of $U(1)_R$ remains unbroken and is 
precisely the MSSM R-parity.
\end{description}

Because the electroweak breaking is mediated by the $SU(2)_L \times 
SU(2)_R$ bidoublet $h$, one expects that the MSSM parameter ${\rm tan} 
\beta \simeq m_t/m_b$.  This has a number of far reaching implications.
Firstly, as discussed in \cite{masses0},
the asymptotic relation $h^{(0)}_b = h^{(0)}_{\tau}$, can be 
exploited to `predict' the top quark mass in the correct mass range, 
$m_t (m_t) \simeq 170$ GeV.  Secondly, by taking the CP-odd scalar mass 
$m_A \stackrel{>}{\sim} m_{Z^{0}}$, the tree level mass of the lightest 
(`Weinberg-Salam') higgs $h^0$ is $m_{Z^0}$.  By including the radiative 
corrections, one finds that $m_{h^0} \approx 105 - 125$ GeV.  Thirdly, 
with ${\rm tan} \beta \simeq m_t/m_b$, and depending on assumptions 
about the SUSY breaking parameters, one can obtain estimates of the 
sparticle spectrum including the LSP.  For instance, with (near) 
universal boundary conditions on the SUSY breaking scalar and gaugino 
masses, one finds that the colored sparticles typically are quite heavy 
($\stackrel{>}{\sim}$ few hundred GeV), one of the staus is the lightest 
charged sparticle, while the LSP is primarily composed from the `bino' 
with mass $\stackrel{>}{\sim}$ 100 GeV \cite{ALS}.

Finally, we note that the breaking of $G_{PS}$ 
to $SU(3)_C \times U(1)_{em}$ gives rise to 
topologically stable superheavy ($\sim 10^{17}$ GeV) monopoles that 
carry two quanta of Dirac magnetic charge \cite{lms}.  
This is consistent with the 
fact that the $G_{PS}$ representations $(1,2,1),(1,1,2)$ and their conjugates 
carry electric charge $\pm e/2$ and are unconfined!  These 'exotics' ,
together with $(4,1,1)$ + $(\bar{4},1,1)$, 
carrying electric charge $\pm e/6$,
belong to the vector-like representations of $G_{PS}$ with masses that may be 
as large as $M_{P}$.  Clearly, some mechanism 
must be found which can suppress the number density of these monopoles 
and their fractionally charged partners to a level below the current 
observational limits. One obvious mechanism would be inflation
which would need to occur beneath the symmetry breaking scale $M$.
In fact, as discussed in DLS \cite{dls}, 
there is an in-built hybrid inflation mechanism
already present in such models due to the fact that the singlet
field $S$ occurs only linearly in the superpotential, and so 
has a very flat potential. For $S$ larger than some critical value
the other VEVs are held at zero and false vacuum inflation may take place,
as $S$ rolls towards the origin. When $S$ reaches some critical value,
the global minimum as discussed here is quickly reached and inflation
ends. Thus there may well be inflation in this model, but it may
not solve the monopole problem if these are produced at the end of
inflation. One possibility is that there is another period of inflation
at a lower scale, as in the next-to-minimal supersymmetric model
of hybrid inflation recently proposed \cite{nmssm}, in which case the
inflation at the higher scale would not seem to be required.
Another possibility is that the symmetry group is taken to be not $G_{PS}$ but
$SU(4) \times SU(2)_L \times U(1)_R$, which does not lead to monopoles.
This would lose the prediction of large $\tan \beta$.

In conclusion, we have presented a straightforward extension of the DLS 
scheme to the gauge group $G_{PS} \equiv SU(4) \times SU(2)_L \times 
SU(2)_R$, and thereby solved the symmetry breaking 
problem for this model. The proposed mechanism also solves the
$\mu$ problem, and provides a mechanism for hybrid inflation.
The symmetry breaking mechanism may be compared to the
simple idea that the squared masses of the $H, \bar{H}$ fields are 
driven negative by some radiative mechanism \cite{LT}.
The equivalent radiative mechanism applied here would be
more indirect, and correspond to replacing the explicit mass
parameter $M$ here by the VEV of a further singlet field,
with the singlet field having its mass squared driven negative
by some radiative breaking mechanism, as recently discussed
by Goldberg \cite{Gold}.
The low energy limit of this model leads to a  number of 
predictions which can be experimentally tested, especially at the LHC.  
What makes this model in particular especially worth exploring is the fact 
that the symmetry $G_{PS}$, in contrast to $SO(10)$, readily arise from 
superstring constructions.  The experimental discovery of `doubly' 
charged monopoles and/or charge $\pm e/2$ color singlets would perhaps 
be the most striking confirmation of the existence of $G_{PS}$ symmetry in 
nature.

\begin{center}
{\bf Acknowledgements}
\end{center}
We are grateful to the Aspen Center for Physics where this work was
started.
We would also like to thank G. Dvali
for useful discussions. 
Q.S. acknowledges the DOE support under grant no.DE-FG02-91ER40626.

\end{document}